\documentclass{aa}       % New LaTeX A&A Standard Fonts
\usepackage{psfig,times}
\renewcommand\deg{\ifmmode^\circ\else$^\circ$\fi}
{\catcode`\@=11
\gdef\SchlangeUnter#1#2{\lower2pt\vbox{\baselineskip 0pt \lineskip0pt
  \ialign{$\m@th#1\hfil##\hfil$\crcr#2\crcr\sim\crcr}}}
}

\def\lessim{\mathrel{\mathpalette\SchlangeUnter<}}

\def\cers#1.{{in: {\it Compact Extragalactic Radio Sources}, ed. J.A. Zensus
    and K.I. Kellermann (NRAO, Socorro), p.~#1.}}
\def\irvine#1.{{in: {\it Quasars and Active Galactic Nuclei: High-Resolution Radio Imaging}, ed.\ M.H. Cohen
    and K.I. Kellermann, {\it Proc.\ Nat.\ Acad.\ Sci.\ USA}, Vol.\ 92, No.\ 5, p.~#1.}}
\def\gammawork#1.{{in: {Proceedings of the Heidelberg Workshop on Gamma-Ray Emitting AGN}, ed.\
    J.G. Kirk, M. Camenzind, C. von Montigny, \& S. Wagner, (MPI-preprint MPI 
    H - V37 - 1996: Heidelberg),  p.~#1. \\
    http://www.lsw.uni-heidelberg.de/projects/extragalactic}}

\begin{document} 
   \thesaurus{ 
   11         % A&A Section 11: galaxies 
               (11.14.1;  % galaxies, nuclei
		11.17.3;  % quasars, general
		11.17.4;  % quasars, individual  
		11.10.1)  % galaxies, jets 
             } 
   \title{Gamma-ray to radio activity and ejection of a VLBI component in the 
	  jet of the S5-quasar 0836+710}
   
  \author{  
    K.\ Otterbein\inst{1,3}  \and 
    T.\ P.\ Krichbaum\inst{1} \and 
    A.\ Kraus\inst{1} \and 
    A.\ P.\ Lobanov\inst{1} \and
    A.\ Witzel\inst{1} \and 
    S.\ J.\ Wagner\inst{2} \and
    J.\ A.\ Zensus\inst{1} \\  
} 
  
  \offprints{K. Otterbein} 
  
\institute{Max-Planck-Institut f\"ur Radioastronomie, Auf dem H\"ugel 69, 
	   D-53121 Bonn, Germany 
           \and
           Landessternwarte Heidelberg, K\"onigstuhl, D-69117 Heidelberg, 
	   Germany 
           \and 
           Now at: Landessternwarte Heidelberg, K\"onigstuhl, D-69117 
	   Heidelberg, Germany
} 
  
   \date{Received ; accepted } 
   
   \maketitle 

%%%%%%%%%%%%%%%%%%%%%%%%%%%%%%%%%%%%%%%%%%%%%%%%%%%%%%%%%%%%%%%%%%%%%%%%%%%%%%%%
  
\begin{abstract}
Broad-band (gamma to radio) variations of the flux density were observed in
the first half of 1992 in the luminous high redshift ($z=2.172$) quasar S5 
\object{0836+710}. VLBI monitoring observations during 1993 -- 1996 performed 
at 86\,GHz, 22\,GHz, 15\,GHz, and 8\,GHz show the ejection of a new jet 
component, which most probably is directly related to a quasi simultaneous 
gamma--, X-ray, optical flaring activity which was observed in February 1992. 
During the period 1992 -- 1993 the flaring propagated through the radio 
spectrum. From several quasi-simultaneous radio spectra taken during this phase 
of activity, we determine the time evolution of the spectral turnover of the 
radio spectrum in the $S_{\rm m}$-$\nu_{\rm m}$ diagram. The data indicate a 
correlation of the jet activity with the variability of the broad-band 
electromagnetic spectrum of the source. The observational findings are 
discussed in the framework of relativistic shock models.

\keywords{galaxies: nuclei -- quasars: general --
	  quasars: S5 0836+710 --  galaxies: jets }  

\end{abstract}

\section{Introduction \label{sc:intro}} 

In early 1992, the ultraluminous S5 quasar \object{0836+710} 
(\object{4C\,71.07}, $z=2.172$) underwent a prominent optical outburst (von 
Linde et al.\ 1993), which later led to enhanced variability in 
the mm- and cm- radio bands
(\cite{mar94}). At the time of the optical flaring, the source also
was in a bright state in
the gamma-regime (EGRET: Fichtel et al.\ 1994) and at soft X-rays
(ROSAT: \cite{bru94}). Motivated by the increased activity of the
source, we started a high frequency VLBI monitoring program in 1993 in
order to investigate the possible relation between broad-band changes in the
flux density and structural variations in the jet on
sub--milliarcsecond scales.  The ejection of new VLBI jet components after
major outbursts in the flux density has been observed also in a few other 
sources (e.\ g.\  component ejection after optical flaring: \object{3C\,345} 
Babadzhanyants and Belokon 1986, 
\object{3C\,273} \cite{kri90a}; ejection after radio flares: 
\object{BL\,Lac} Mutel et al.\ 1990, \object{PKS\,0420$-$014}, 
Wagner et al.\ 1995a) and seems to be a quite common phenomenon 
in compact flat spectrum radio sources (blazars). 

With the new data obtained for \object{0836+710}, it appears very likely
that the relation between outbursts in the flux density observed at high 
frequencies and jet component ejection is not limited to the synchrotron regime (e.g. optical-to-radio), 
but most probably covers a much wider spectral range (gamma-to-radio), 
possibly the full electromagnetic spectrum.
Similar correlations between gamma-activity and jet
activity were recently found also in a number of other sources, e.g. in
3C~273 (Krichbaum et al.\ 1996), 3C~279 (Wehrle et al.\ 1994), 0528+134 and
3C454.3 (Krichbaum et al.\ 1995), where the ejection of jet components
could be related to gamma-flares, or enhanced levels of gamma-activity.
However, in most of the published cases the frequency coverage is
not as broad as presented here for \object{0836+710}.

The quasar \object{0836+710} is a member of a complete flux density limited
sample of 13 flat--spectrum radio sources (Eckart et al.\ 1986 \& 1987)
compiled from the 6\,cm S5 survey (K\"uhr et al.\ 1981). 
The 13 sources of the sample are located
in the northern polar cap ($\delta \geq 70\deg$) and are regularly
studied with VLBI at frequencies ranging from 0.3 to 22\,GHz 
(cf. Witzel et al.\ 1988 and references therein).  The VLBI data for
\object{0836+710} now span a time range of more than 15 years (1983--1996). 

The high dynamic range VLBI images of \object{0836+710} reveal a complex and 
wiggled one--sided core--jet structure (Eckart et al.\ 1986, \cite{eck87}). 
World-array VLBI images at 90\,cm and 18\,cm show the jet extending over 
more than 150\,milliarc\-seconds (mas) in direction to the outer 
arcsecond lobe seen with MERLIN 
and the VLA (\cite{hum92}). Mainly from the VLBI monitoring at 6\,cm
(Krichbaum et al.\ 1990b hereafter K90b, Otterbein 1996 hereafter O96), a very 
complex motion pattern is revealed, with jet components, moving at velocities 
ranging from subluminal ($\beta_{\rm app}=0.5; \beta = v/c$) to superluminal 
($\beta_{\rm app}=10$) (assuming $\rm H_0 = 100\, \rm{km/s/Mpc}$ and $q_0 = 
0.5$, throughout this paper).
No systematic correlation between the speed of the jet and the separation of 
the components from the core seems to be present. The jet shows a noticeably
structure, with lateral
displacements of its ridge line (kinks) and oscillations of its transverse 
width (K90b).

At cm-wavelengths, \object{0836+710} is strongly
polarized ($\rm p \simeq 9\,\%$ at 5\,GHz).  Polarization VLBI images
show a highly polarized jet with the magnetic field roughly following the bent
ridge line (\cite{caw93}). This suggests that the jet of \object{0836+710}
is also highly magnetized.

In this paper, we focus on the structural changes observed in the 
sub-milliarcsecond regions of the inner jet of \object{0836+710} during 
and after the optical outburst of 1992. A more detailed discussion of the
overall morphology and the spectral and kinematic properties of the 
milliarcsecond-jet will be given elsewhere.

\begin{table}

{\small
\caption{\label{sum} Summary of experiments}

\begin{tabular}{lcll}
date            &  freq.   &observing mode,   & array          \\
                &  [GHz]   &bandwidth [MHz]   &                \\ \hline
1993, Apr. 7    &  86.2    &MK\,III (A), 112 & global (4)$^1$  \\
1993, Sep. 16   &  22.2    &MK\,III (B), 56  & global (11)     \\
1995, Jan. 20   &  15.3    &VLBA 64-8-1, 32 & VLBA (10)$^2$    \\
1995, Aug. 24   &  22.2    &VLBA 128-8-1,64 & VLBA (10)$^3$    \\
1995, Aug. 24   &  ~8.4    &VLBA 128-8-1,64 & VLBA (10)$^3$    \\
\end{tabular}

\vspace{0.3cm}
Notes: \\
$^1$: stations: Effelsberg, Pico Veleta, Onsala, Haystack\\
$^2$: snapshot--type observations \\
$^3$: 8/22\,GHz simultaneous dual--frequency observations \\
}
\end{table}

\section{Observations and data reduction \label{sc:data}}
\noindent
\object{0836+710} was observed with VLBI between 1993 and 1995 
at frequencies ranging from 8\,GHz to 86\,GHz. The observations
in August 1995 were made in dual-frequency mode (22 and 8\,GHz). At 86\,GHz and 
15\,GHz\footnote{The 15\,GHz data were kindly provided by A. Patnaik}
the data for \object{0836+710} were obtained during VLBI campaigns
which also included other sources (86 GHz: Schalinski
et al.\ 1994; 15\,GHz: Patnaik et al.\ 1996). Owing to the limited 
$uv$--coverage in the 86 and 15\,GHz data sets, the
dynamic range of the corresponding maps is lower than that 
of the high dynamic range images at 22 and 8 GHz, with their 
optimal uv-coverage.
The details of the observations are summarized in
Table \ref{sum}. Column 1 gives the observing date, column 2 the frequency,
column 3 the observing mode and bandwidth, and column 4 the VLBI array
with the number of participating stations in brackets. 

After the correlation (VLBA data in Socorro, MK\,III data in Bonn),
the data were fringe-fitted and calibrated in the 
standard manner. AIPS was used for reducing the VLBA data. 
The fringe fitting of the MK\,III data was done using the Haystack 
software (Rogers et al.\ 1983),
and the baseline oriented global fringe fitting routines available at
the MPIfR (Alef \& Porcas 1986, Alef 1989). The amplitude calibration was
carried out in the standard way, using system temperature measurements and
elevation-dependent antenna gain curves.
At 22\,GHz and 86\,GHz we included corrections for atmospheric absorption 
(opacity) in the amplitude calibration.

The imaging was performed using the CalTech VLBI package and the difference
mapping software DIFMAP (Pearson 1991, Shepherd et al.\ 1994). In parallel to 
the mapping, we made Gaussian model fits to the visibility
amplitudes and phases, in order to parameterize the 
multi--component structures seen in the maps. The measurement errors 
of the parameters of the Gaussian
components were determined formally from least square fits and, in
addition, from their variations on fits to slightly different calibrated and 
edited data sets. 

\section{Results \label{sc:results}} 

\begin{figure}
\mbox{\psfig{figure=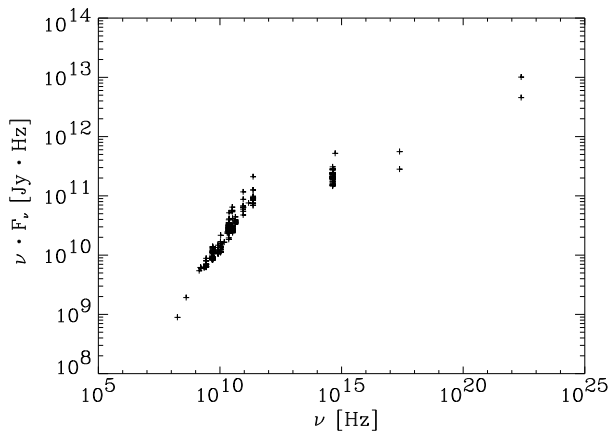,width=7cm,width=8.5cm,angle=90}}
 \caption[]{\label{fg:nufnu}
          Overall spectrum ($\rm \nu F_{\nu}$) of \object{0836+710} as compiled
          from presented and published data.}
\end{figure}

\subsection{Broad-band Variability \label{sc:bband-var}}

\begin{figure} 
 \begin{center}
  \mbox{\psfig{figure=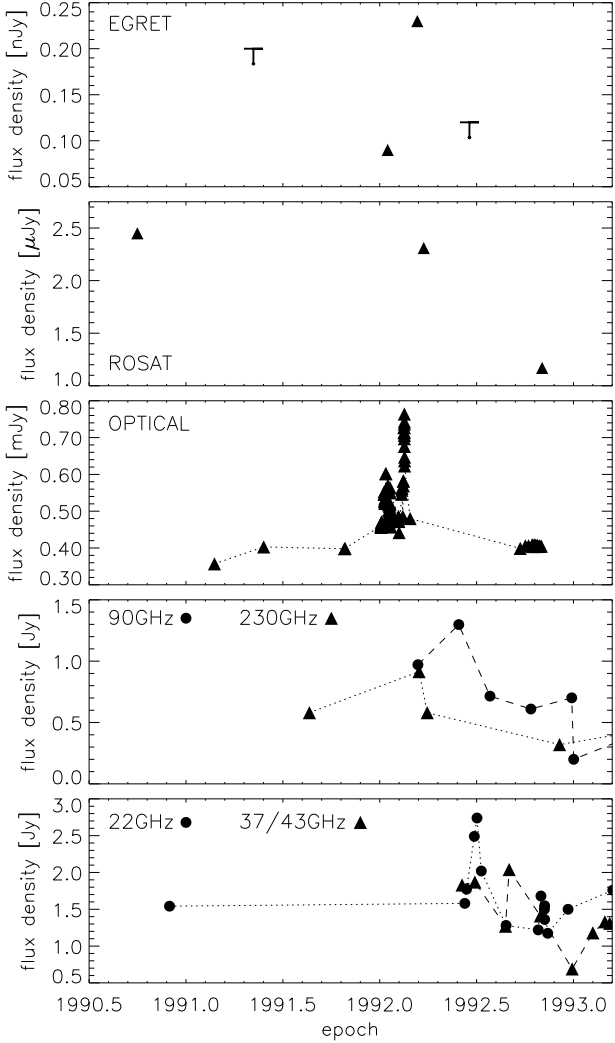,width=9cm}}
\vspace*{0.4cm}
\caption[]{\label{fg:lightcurves} Broad-band flux density variability in 
	   \object{0836+710} for the period 1990.5 -- 1993.2. From top to 
           bottom gamma-ray, X-ray, optical, millimeter- and cm--radio data
           are shown (for origin of data see text). The lines connect adjacent
           data points, arrows in the top panel indicate upper limits.
           The data suggest a correlated gamma-- to optical outburst first peaking
           in 1992.13, and propagating later towards longer (\mbox{mm-/}cm-) wavelengths.}
 \end{center}
\end{figure}

\object{0836+710}, which in the past
showed only mild variations of its radio flux density,
recently changed into a more active phase of variability.
Between 1990--1993 the source exhibited strong variability,
which seems to be correlated over a wide range of the electromagnetic 
spectrum. In Figure \ref{fg:nufnu}, we show the total luminosity
spectrum; and in Figure \ref{fg:lightcurves}, we show the flux density 
measurements for the interval 1990.5--1993.2. The data are taken from the 
literature and include, from the top to the bottom, EGRET observations in the 
gamma-band at 100\,MeV (Mukherjee et al.\ 1997), ROSAT observations in the 
X-ray band at 1\,keV (Brunner et al.\ 1994), optical R-band observations (von 
Linde et al.\ 1993, Schramm et al.\ 1994 and Heidelberg telescope), 230\,GHz 
and 90\,GHz flux density measurements from the IRAM 30\,m millimeter telescope 
at Pico Veleta and from the JCMT at Mauna Kea (Steppe et al.\ 1993, Bloom et 
al.\ 1994), and measurements at 43, 37 and 22\,GHz from the Mets\"ahovi flux 
density monitoring program (E.\ Valtaoja, priv.\ comm.) and from the 100\,m 
radio telescope at Effelsberg.

From Figure \ref{fg:lightcurves}, it is clear that the sparse
and irregular time sampling, particularly in the X-ray and
gamma-bands, requires a careful interpretation. Obviously,
a cross correlation analysis, or an accurate determination of the time lags 
between the different bands is not possible. The data nevertheless 
indicate a tight correlation between the gamma--/X-ray and optical flux 
density levels. Close to the optical flaring activity observed in 
1992.1--1992.2, the gamma-- and X-ray flux density was also high. The EGRET 
measurements show variability by a factor of 2--3 on a time scale of 
$< 0.5$\,yrs, and low flux densities shortly before and after the time of the 
optical flaring. The X-ray flux is also variable. About 0.5 years after the 
optical outburst, it dropped by a factor of 2. Given the sparse sampling,
the data are consistent with a (at least nearly) simultaneous 
($\Delta t \leq 0.1$\,yrs) gamma-ray/optical flaring. This is consistent
with nearly simultaneous gamma--/optical variations observed in a few
other objects, eg. in 1406$-$076 (Wagner et al. 1995b).
From the measurements in the mm- and cm-bands it appears that the optical
outburst propagates towards longer wavelengths, with the maximum in the
radio appearing after $0.1-0.3$\,yrs first at shorter (mm-), and
after $0.3-0.5$\,yrs later at longer (cm-) wavelengths.

\subsection{Results from the VLBI monitoring \label{sc:vlbi-mon}}

\subsubsection{Overall morphology and component identification}
\begin{figure*}
\begin{center}
\mbox{\psfig{figure=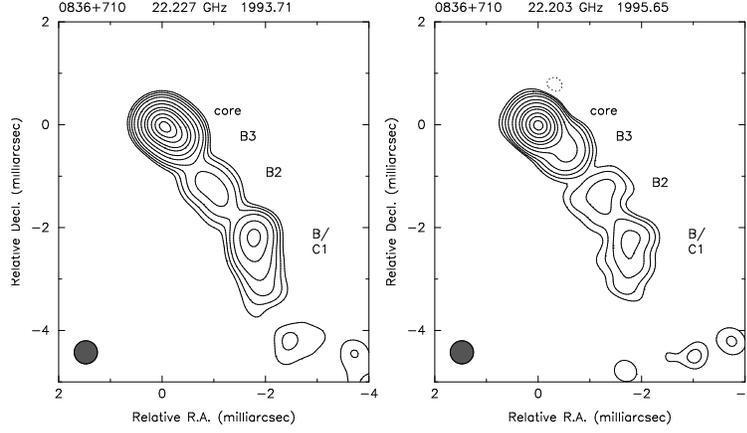,width=10cm,angle=270}}
 \caption[]{\label{innerjet} 
	   22\,GHz maps of the inner jet of \object{0836+710} from 
	   1993.71 (left) and 1995.65 (right). Contour levels are 
	   --0.25, 0.25, 0.5, 1, 2, 5, 10, 20, 30, 50, 70, 90\,\% of the peak 
	   flux density of 0.35 Jy/beam (left), and 0.9 Jy/beam (right). Both 
	   maps are restored with a circular beam of 0.45\,mas size.
           Labels denote the component identification.}
\end{center}
\end{figure*}

In Figure \ref{innerjet}, we show two 22\,GHz images of \object{0836+710}
obtained in 1993.71 and 1995.65. As in the maps obtained earlier
at longer wavelengths, the source shows a one sided core/jet
structure with several compact components. From the maps and the model
fits, we identified several distinct jet components B3, B2, B, and C1 (B
and C1 are blended) located at separations from the core of up to $r \simeq 
4$\,mas. The labels in Figure \ref{innerjet} 
reflect this identification 
scheme, in which we
implicitly assumed that the brightest component located at the map
center is the stationary core of the jet. Beyond a separation of 4~mas
from the core, the jet emission becomes weaker and more extended, so that it
is resolved by the interferometer beam. Images obtained by applying a taper to
the 22~GHz $uv-$data, however, show that the emission from the jet extends to 
15~mas from the core, in good agreement with the outer structure seen 
previously at longer wavelengths (K90b, Hummel et al.\ 1992).

In the following, we use a component annotation similar to the one introduced
by K90b for the 5\,GHz data. The feature seen at a separation of 
$r \sim 3$\,mas from the core can be identified as a blend between a fast 
moving older 5\,GHz component B ($\beta_{\rm app}=9.6 \pm 0.2; \beta= v/c$) and 
a quasi-stationary component C1 ($\beta_{\rm app} \leq 2$). Near component
C1, the jet direction changes. This displacement is seen more
clearly in the 'tapered' 22~GHz images, and also 
in the maps at 15 and 8
GHz. We note that similar displacements also appear at larger separations from
the core, in particular near component D at $r \simeq 10$\,mas in
the 5 GHz maps (K90b), and near component G2 at
$r \simeq 30$\,mas (Hummel et al.\ 1992) at 1.6\,GHz. With the
new 22\,GHz maps, it appears very likely that the lateral displacements 
(kinks) of the jet axis are systematic and reflect a general jet property in 
\object{0836+710}.

At smaller separations from the core, two new components B2 \& B3 are seen
(Figure \ref{innerjet}). Linear back--extrapolation of the motion of
B2 allows its identification with a component previously visible in a 5\,GHz map
of 1990 (O96). Adopting its constant apparent speed of
$\beta_{\rm app}=8.3 \pm 2$, the component B2 was ejected before 1986.8 and
therefore cannot be related to the flare observed in 1992.

\begin{figure}
\mbox{\psfig{figure=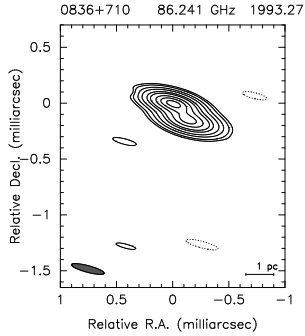,width=7cm,angle=270}}
\caption[]{\label{3mmmap} 
	    86\,GHz map of \object{0836+710} of 1993.27. Contour levels are 
	    --1, 1, 2, 5, 10, 20, 30, 50, 70, 90\,\% of the peak flux density 
	    of 0.41 Jy/beam. The observing beam size is $0.30 \times 
	    0.06$\,mas, $\rm{P.A.} = 75^\circ$.}
\end{figure}

In the 22\,GHz map of 1993.71, the core appears elongated, indicating
the emergence of a new component (see Fig.\ \ref{innerjet}, left). From 
Gaussian model fits we determined its relative separation from the core of 
$r \simeq 0.3$\,mas. We label this component B3.  The image of 1995 
(Fig.\ \ref{innerjet}, right) shows B3 at $r \simeq 0.8$\,mas, thus 
indicating superluminal motion with a speed similar to its predecessors B2 and 
B (see next paragraph).

The ejection of a new component and its subsequent separation from the core is 
further supported by mm-VLBI observation of the source. A high angular 
resolution map obtained at 86~GHz in 1993.27 (Fig.\ \ref{3mmmap}), 5 months 
earlier than our first 22\,GHz observation, showed a secondary component
next to the unresolved VLBI core (size  $\leq 0.1$\,mas). Owing to the small 
number of participating VLBI stations at this high observing frequency and
the resulting limitations of the $uv$-coverage, the 
86~GHz map shows only the most prominent features of a probably more complex 
brightness distribution. The secondary component is located at $r \simeq 
0.15$\,mas. 
Since its position agrees well with the extrapolated position of B3 (from 22 GHz 
data) at that time, we identify this component with B3.
 
\subsubsection{Motion of component B3 \label{sc:B3}}

In Table \ref{rvst}, we summarize the relative separations of B3 from the core
for all available observing epochs. We further added a data point from
a 22~GHz map of 1996.34, which was kindly provided by A.\
Marscher (priv.\ comm.).

\begin{table} 
\caption{\label{rvst} Relative separation from the core of the new component B3}
\begin{tabular}{lccc}
epoch &freq. &  R            &     PA   \\
      &[GHz] &[mas]          & [$^\circ$]   \\ \hline
1993.27&86& 0.15 $\pm$ 0.05  & 220 $\pm$ 20  \\
1993.71&22& 0.34 $\pm$ 0.04  & 231 $\pm$ 3  \\
1995.05&15& 0.72 $\pm$ 0.15  & 227 $\pm$ 5  \\
1995.65&~8& 0.69 $\pm$ 0.15  & 229 $\pm$ 5  \\
1995.65&22& 0.76 $\pm$ 0.06  & 230 $\pm$ 3  \\
1996.34$^1$&22& 1.05 $\pm$ 0.10  &  --          \\
\end{tabular}

\vspace{0.3cm}
Notes:$^1$: A. Marscher, priv. comm.\\
\end{table}

\begin{figure}
 \mbox{\psfig{figure=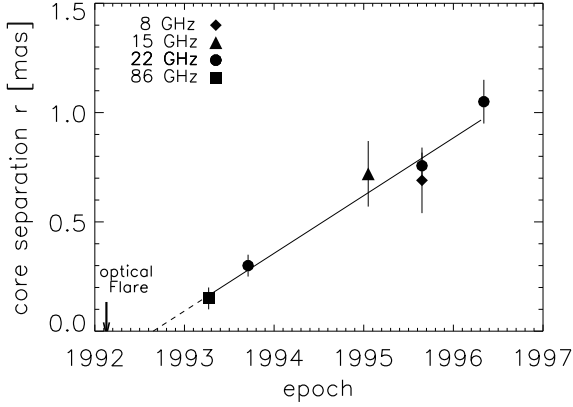,width=8.5cm,angle=90}}
 \caption[]{\label{birth} Relative separation of component B3 from the core 
	     plotted versus time. Symbols denote different frequencies:
             diamonds (8~GHz), triangles (15~GHz), circles (22~GHz), and squares
	     (86~GHz). The solid line represents a linear fit with slope $\rm 
	     \mu =0.26$ mas/yr, corresponding to superluminal motion with 
	     $\beta_{\rm app} = 10.8$. Linear back-extrapolation (dashed line) 
	     yields component ejection near 1992.65.
             The arrow indicates the onset of the optical flare 
	     (1992.13).}
\end{figure}

In synchrotron self--absorbed VLBI jets (e.g. Blandford and K\"onigl
1979), frequency dependent shifts of the relative positions of the
VLBI components, caused by optical depth effects, are expected. The
comparison of the relative component positions between the nearly
simultaneous 8, 15 and 22~GHz images, however, indicate that the
position shifts in \object{0836+710} are relatively small ($\lessim
0.1$~mas). In the following, we neglect such shifts due to opacity 
and plot in Figure \ref{birth} the 
relative separation from the core of B3 versus time. The data points line up
very nicely, thus confirming indirectly the validity of our assumption
of negligible opacity effects. Furthermore, we find no evidence for
non-linear motion, e.g. acceleration or motion along curved paths
(see also the nearly constant position angle of B3 in table \ref{rvst}).

A linear fit to the data yields an apparent angular separation rate of
$\mu = 0.26 \pm 0.03$~mas/yr corresponding to an apparent velocity of
$\beta_{\rm app}= 10.8 \pm 1.3$. This velocity is close to the speeds seen for 
the older components B \& B2 (K90b, O96),
and confirms the previous finding of relatively high velocities at small 
separations from the core (note: components C1 and C2 move substantially
slower, cf. K90b). Linear back-extrapolation of the motion of B3 yields the 
time of zero separation of $t_{\rm ej}=1992.65^{+0.15}_{-0.17}$. This is close 
to the time of the strong optical flare observed in 1992.13 (von Linde et
al.\ 1993, see Figure \ref{fg:lightcurves}). Within possible
extrapolation errors and due to the limited and partially incomplete time coverage 
of the optical flaring, it therefore appears that B3 was
ejected {\it during or shortly after} ($0 \lessim \Delta t \leq
0.67$\,yrs) the optical outburst. We study the correlation between
the optical flaring and the ejection time of B3 in more detail in section 
\ref{sc:travel}.

\subsection{Evolution of the total radio spectrum \label{sc:sp-evol}}

\begin{figure}[t]
\mbox{\psfig{figure=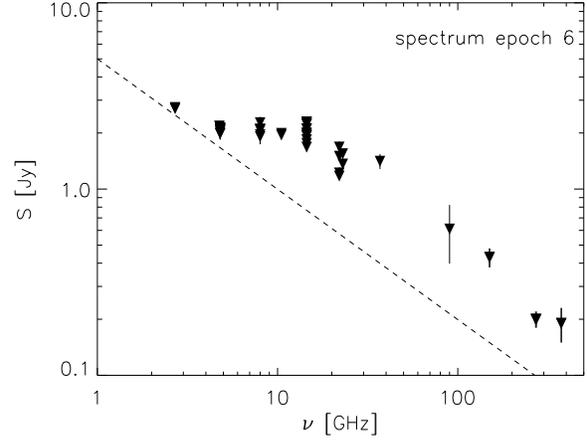,height=6.5cm,width=8.5cm,angle=90}}
\mbox{\psfig{figure=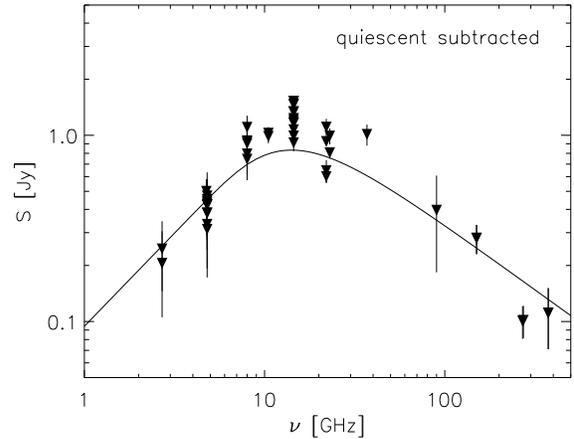,height=6.5cm,width=8.5cm,angle=90}}
\caption[]{\label{fg:spectrum} 
           Spectrum of \object{0836+710} at 
           epoch 6 (1992.75). In the upper picture the total 
           spectrum is plotted. The dashed line shows the quiescent
           spectrum, i.e.\ the quiescent part determined from an observations
           prior to the optical flare in 1992. In the lower panel
           the residual variable part of the spectrum (quiescent spectrum 
	   subtracted) and the fit used to determine the spectral turn over is 
	   shown.}
\end{figure}

Figure \ref{fg:lightcurves} shows the propagation of the outburst
through the spectrum. Such a behavior is readily
described within the standard shock--in--jet models (e.g. Marscher \&
Gear 1985), in which a relativistic shock propagates along a synchrotron 
self-absorbed jet, causing (at least qualitatively) the observed 
frequency dependence.

From the data shown in Figure \ref{fg:lightcurves} and additional data from
measurements made with the MPIfR 100\,m radio telescope at longer
cm--wavelengths, we are able to follow the evolution of the
radio spectrum and to determine the path of a local spectral
excess, which propagated from higher to lower frequencies.

From our radio data, we first constructed quasi-simultaneous radio spectra 
covering the $1.4-230$\,GHz range.  
Good spectral coverage is achieved in the
period from 1992.05 to 1993.00.  To overcome the lack of frequency coverage
at individual epochs, we binned the available data into 7 intervals,
each of 50 days in duration.  
Since the source did not show strong variations within the binning-timescale,
we are confident that the averaging of the data within each bin
is not causing large errors for our later derived turnover points.
Figure \ref{fg:spectrum} (top) gives an example of such a radio spectrum.

The long-term flux density monitoring of \object{0836+710} at cm-wavelengths
shows longer periods of low or only moderately slow (timescales of
$\sim$years) variability. One such period had occurred prior to the time of
the optical activity in 1992. We used the spectra from this period
to determine the shape of the quiescent spectrum (non-variable part). 
Throughout the radio band considered ($\sim$ 1--100\,GHz) it can be well 
represented by a simple power law ($S_\nu \propto \nu^{-0.7}$), which we plot 
as dashed line in Figure \ref{fg:spectrum} (top).

\begin{figure}[t]
\mbox{\psfig{figure=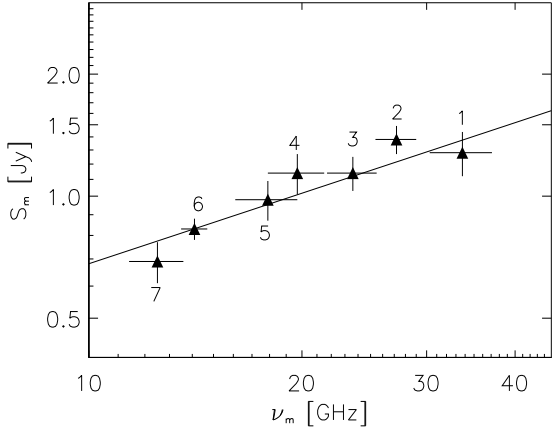,height=6.5cm,width=8.5cm,angle=90}}
\caption[]{\label{fg:sm_num} 
          Path of the spectral evolution of the turnover point in the
          $S_{\rm m}$--$\nu_{\rm m}$-diagram. The labels 1-7 denote the individual 
	  spectral epochs
          (see text). The straight line shows a power law fit 
          ($S_{\rm m} \propto \nu_{\rm m}^{0.58}$) to the data.
	  }
\end{figure}

In order to determine the position ($\nu _{\rm m}$, $S_{\rm m}$) of the turnover
point of the variable part of the radio spectrum, we now subtracted the 
same quiescent spectrum from our binned spectra. The fit of a
homogeneous synchrotron component spectrum (with spectral indices
$\alpha_{\rm thick}=2.5$ and $\alpha_{\rm thin}=-0.7$) to each of the
residual spectra was rather unsatisfactory, indicating that the
variable part of the radio spectrum cannot be represented well enough by a
single and homogeneous synchrotron component. A much better
fit was achieved when we used $\alpha_{\rm thick} = 1.0$ and 
$\alpha_{\rm thin} = -0.7$. In the following we use the results from these fits.
Figure \ref{fg:spectrum} (bottom) shows an example of such a fit for one of the 
spectra (epoch 6). From each of the fits to the 7 
residual spectra, we determined the position of the spectral turnover. 
In Figure \ref{fg:sm_num} the turnover flux density $S_{\rm m}$ is plotted
versus turnover frequency $\nu_{\rm m}$ as a function of  time.
The numbered labels at each position correspond to the following epochs: 
1 - 1992.07 (TJD 8650), 2 - 1992.21 (TJD 8700), 3 - 1992.35 (TJD 8750),
4 - 1992.48 (TJD 8800),
5 - 1992.62 (TJD 8850), 6 - 1992.76 (TJD 8900), 7 - 1992.90 (TJD 8950). 

The measured path of the spectral turnover seems to
follow a linear trend (in the $log(S_{\rm m})/log(\nu_{\rm m})$ space), which 
can be interpreted within models invoking a shock moving along a relativistic 
synchrotron self--absorbed jet. In the model of Marscher \& Gear
(1985) and Marscher (1990), the spectral evolution is described
by three subsequent phases of energy loss and radiation,
the Compton, synchrotron, and adiabatic
phase. During the first two stages, the
turnover flux rises with decreasing frequency, then remains nearly constant.  
The phase of adiabatic losses is characterized by a decrease of the
flux density with decreasing turnover frequency. We therefore identify the
spectral evolution seen in Figure \ref{fg:sm_num} with the adiabatic--loss
stage. A power law fit (the
straight line in Figure \ref{fg:sm_num}) to the path of the turnover
point yields $S_{\rm m} \propto \nu_{\rm m}^{0.58}$.

We note that in Figure \ref{fg:sm_num} a slightly steeper slope is
obtained, if the first data point (point no. 1) is not included in the 
fit. Within the afore mentioned model, it is possible that
the spectral evolution from point no. 1 to no. 2 reflects the transition
from the synchrotron to the adiabatic cooling phase. If this is the case,
the synchrotron loss dominated phase would have ceased $t_{\rm sync} < 30$\,days
after the onset of the optical flare. Obviously observations much closer to the time
of the outburst would have been required to study the synchrotron phase.

\section{Discussion \label{sc:disc}}

\subsection{The total spectrum and the VLBI jet \label{sc:model}}

The $S_{\rm m}$--$\nu_{\rm m}$ dependence determined in the previous section can
be used for studying the relation between the spectral and kinematic
properties of the relativistic jet in \object{0836+710}. Despite some
evidence for emission from an inhomogeneous synchrotron component
($\alpha_{\rm thick} = 1.0$, see previous section), we postulate for
the sake of simplicity, that the observed spectral variations are produced
by a single, compact relativistic shock dominating the source radio
emission, and associated with the core of the jet or an emitting region
moving inside the jet within $\lessim 0.1$\,mas distance from the core
(a distance comparable to the back--extrapolated separation of B3 at
the last spectral epoch, 1992.9). We use the formalism developed by
Marscher \& Gear (1985), and account for possible variations of the
Doppler factor in the emitting region (Marscher 1990). Based on the
synchrotron spectral index $\alpha=-0.7$ used for the spectral
fitting, a power law electron energy distribution with $s=2.4\,\,
(s=1-2\alpha)$ is assumed. The magnetic field distribution along the jet axis 
$r$ is defined by an exponent $a$, so that $B \propto r^{-a}$.  The Doppler
factor is $\delta \propto r^b$. The spectral evolution of the shock is
then described by the variations of the turnover point: $S_{\rm m}\propto
\nu_{\rm m}^{\rho}$ and $\nu_{\rm m} \propto r^{\epsilon}$.  The exponents $\rho$
and $\epsilon$ depend on the dominating type of the energy losses
(Marscher 1990, Marscher et al.\ 1991, Lobanov \& Zensus
1998). Following Marscher (1990), we obtain the following expressions
for $\rho$ and $\epsilon$ during the adiabatic--loss stage:
\begin{equation}
\label{eq:index_Sm}
\rho = -\frac{(19-4s) - 3a(2s+3) + 3b(3s+7)}{2(2s+1)+3(a-b)(s+2)}\,,
\end{equation}
\begin{equation}
\label{eq:index_num}
\epsilon = -\frac{2(2s+1)+3(a-b)(s+2)}{3(s+4)}\,.
\end{equation}

We now consider two cases: a)~a jet
with transversal ($a=1$) magnetic field, and b)~a jet with longitudinal 
($a=2$) magnetic field. With the measured $\rho=0.58$, changes of the Doppler
factor and turnover frequency along the jet can be determined from 
(\ref{eq:index_Sm}) and (\ref{eq:index_num}). 

For $a=1$, we obtain $b=0.0$ and $\epsilon=-1.3$. The distance
travelled by the emitting region between 1992.07 and 1992.90 (the 1st
and the 7th spectral epochs) should increase by a factor of
$r_7/r_1\approx 2.1$. To assess the jet kinematics, we use
$\beta_{\rm app} = 10.8$ measured in Section \ref{sc:B3} for the jet component
B3 at small separations from the core. For now, we assume that the jet bulk
Lorentz factor is $\gamma_{\rm j}= 10.9$, with the
resulting Doppler factor $\delta_{\rm j} = 11.4$ and jet viewing 
angle $\theta_{\rm j}=5\deg$. Using the time separation between the two
aforementioned spectral epochs, $\Delta t = 0.83$\,yrs, one can then
estimate the distance between the emitting region and the jet apex,
$r_1 = \beta  \gamma \delta \Delta t / (1+z) = 9.5\pm0.8$\,pc.

The case with $a=2$ results in $b=0.4$, implying that the emitting region
was accelerating. The corresponding turnover frequency evolution is
described by $\epsilon=-1.7$. The resulting Doppler factor must increase
between $t_1$ and $t_7$ by $\delta_7/\delta_1 = 
(\nu_{\rm m\,7}/\nu_{\rm m\,1})^{b/\epsilon} \approx 1.3$, and the corresponding 
distance must become larger by a factor of $\approx 1.8$.
For a jet with variable Doppler factor, the distance travelled by the
shock between two
epochs $t_1$ and $t_2$ is (Lobanov \& Zensus 1998):
\begin{equation}
\label{eq:travdistance}
\Delta r_{1,2} = (1+z)^{-1} \int_{t_1}^{t_2} \frac{\beta(t) {\rm d}t}
{1 - \beta(t) \cos\theta(t)}\,.
\end{equation}
For the Doppler factor $\delta \propto r^b$, equation (\ref{eq:travdistance})
gives for the distance to the jet apex at the epoch $t_1$:
\begin{equation}
\label{eq:zerodistance}
r_1 = \left( \frac{1+z}{\delta_1 c \Delta t} \int_{1}^{r_u} 
\frac{1}{\sqrt{\gamma^2(r)-1}} \frac{{\rm d}r}{r^b} \right)^{1/(b-1)}\,,
\end{equation}
where $c$ is the light speed, $\Delta t = t_2 - t_1$, and $r_u =
(\nu_{\rm m\,2}/\nu_{\rm m\,1})^{1/\epsilon}$. The exact form of
$\gamma(r)$ is essentially unknown. In many cases, it can be assumed
to be constant (so that the Doppler factor variations are entirely due
to a curved path followed by the emitting region). For a straight jet,
the $\gamma(r)$ can be derived from the Doppler factor variations. For
\object{0836+710}, both approaches yield similar result: $r_1 \sim 30$\,pc 
(here we postulate $\delta_7 = \delta_{\rm j}$). One
can argue that the location $r_1$ (the beginning of the
adiabatic--loss stage) should correspond to the VLBI core, since the
core emission is likely to be dominated by Compton and synchrotron
energy losses (Unwin et al.\ 1994, 1997), whereas the moving jet
components are successfully modelled as adiabatically expanding
spherical plasmons embedded in the jet (Zensus et al.\ 1995). In view
of this argument, the case of a non--accelerating jet with $r_1\approx
10$\,pc appears to be more likely to explain the observed spectral
behavior in \object{0836+710}.

\subsection{Travel times and radio--to--gamma correlations \label{sc:travel}}

If the jet core is located at $r_1$, one can calculate the travel time
between the jet nozzle, $r_0$, and the core. For $b=0$
($\delta=$const), the calculation is trivial, and it gives (in the
observer's frame) $\Delta t_{0,1} = (1+z) r_1 \gamma^{-1} \delta^{-1}
\beta^{-1} = 0.74\pm0.07$\,yrs.  For an accelerating jet, the travel
time can be calculated from (\ref{eq:travdistance}), and it depends on
the (unknown) Doppler factor $\delta_0$ at the jet nozzle. If the
nozzle is formed by a pressure drop in a relativistic outflow,
the resulting Lorentz factor is $\gamma_0 \approx 1.2$ (Marscher
1980). The corresponding Doppler factor is then $\delta_0 \approx
1.8$, for $\theta_0 = \theta_{\rm j} = 5\deg$. The resulting travel
time is $\Delta t_{0,1} = 1.6$\,yrs. For larger $\gamma_0$, the travel
time increases and may become as large as 2.8\,yrs, if $\gamma_0
\approx \gamma_1$.

On the basis of the arguments presented in the previous section, we take 
$\Delta t_{0,1}  = 0.74$\,yrs as a better estimate of the travel time 
between the nozzle and the core. In section \ref{sc:B3}, we have found 
$t_{\rm ej}=1992.65$ for
the ejection time of the superluminal feature B3. Together with the derived
nozzle--to--core travel time, this implies that the plasma condensation
responsible for the emission of B3 had travelled through the nozzle at 
$t_0 = 1991.91\pm0.21$. The latter epoch is very close to the time
when the optical flare was observed in the source. Although the coincidence 
may be only fortuitous, it is rather striking, and we would like to emphasize 
that such a situation is indeed possible. For instance, if a 
flare occurs when a dense plasma condensation travels through the
jet sonic point, the optical depth in the radio bands can exceed unity, 
and only a high--energy flare (optical, X--rays, or gamma--rays) is observed. 
An increase of the radio emission and the associated ejection of a new 
VLBI feature are observed later, after the condensation has travelled
outside the $\tau\geq 1$ region.

The jet geometry and travel times above are determined for $
\gamma_{\rm j}\approx\gamma_{\rm min}$. If we suppose that the gamma-to-optical
flaring activity and the ejection of VLBI component B3 in 1992 had a common
origin, the jet geometry can be uniquely determined. We assume that the maximum
of the flare 
occurs at $t_{\rm flare} = 1992.13$ and take $t_{\rm ej} = 1992.65$ from the 
extrapolation of the component motion. In such a
scenario, $\Delta t_{0,1} = t_{\rm ej} - t_{\rm flare}$, and we find
$\gamma_{\rm j} = 11.8$ and $\theta_{\rm j} = 3\fdg2$ . The corresponding
Doppler factor is $\delta_{\rm j} = 16.4$; the core--to--nozzle distance
is $r_1 \approx 15$\,pc. We note that the derived values are
consistent with overall kinematics of \object{0836+710} (O96).  We
therefore view this model as a likely explanation for the
kinematic properties, spectral behavior and flaring activity observed
in \object{0836+710} in 1992--1993.

\subsection{Properties of the inner jet \label{sc:inner-jet}}

The kinematic parameters determined above for the case of a
non--accelerating ($b=0$) jet with transverse ($a=1$) magnetic field
can be further used to evaluate the physical conditions in the
emitting region responsible for the observed spectral changes in
\object{0836+710}. For this purpose, we associate the core--to--nozzle 
distance, $r_1 = 15$\,pc with the location of the emitting region at
the estimated epoch of origin of B3 ($t_{\rm ej} = 1992.65$). Then, for
each spectral epoch, we find the respective location,
\begin{equation}
r = r_1 \left(\frac{\nu_{\rm m}}{\nu_1}\right)^{1/\epsilon}\, ,
\label{eq:jet5}
\end{equation}
and size of the emitting region (using equation (3) from Marscher 1987),
\begin{equation}
d = d_1 \left(\frac{\nu_1}{\nu_{\rm m}}\right)^{(5\epsilon + a+b )/4\epsilon}
\left(\frac{S_{\rm m}}{S_1}\right)^{1/2} \, .
\label{eq:jet6}
\end{equation}
$S_1$ and $\nu_1$ are measured at $t_{\rm ej}$, which gives $\nu_1 =
17.9\pm1.8$\,GHz and $S_1 = 0.98\pm0.11$\,Jy. For $d_1$, we use the
size of the VLBI core, $\Omega_{\rm core}\approx 0.1$\,mas, typically
measured in the 22\,GHz images of \object{0836+710}. The corresponding
linear size is $d_1 = 0.4$\,pc.

\begin{figure}[t]
\mbox{\psfig{figure=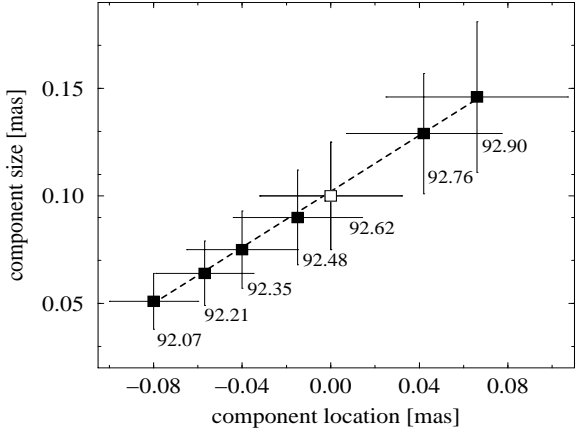,height=6.5cm,width=8.5cm,angle=0}}
\caption[]{\label{fg:jetexp} 
	   Projected location and size of the emitting feature as 
           derived from the observed spectral evolution. The locations of 
           the emitting feature are calculated assuming a viewing angle of 
           $\theta_{\rm j}=3\fdg2$ obtained from equating the core-to-nozzle 
           travel time with the observed time-lag between the gamma-flare 
           and the ejection of B3. The respective spectral epochs 1--7 are
           given next to the data points. Open square denotes the location 
           of the VLBI-core of \object{0836+710}. The increase of the size 
           of the emitting region can be modeled by a conical jet 
           with an opening angle $\phi_{\rm j} = 2\fdg1\pm0\fdg1$.
	  }
\end{figure}

The derived projected location and size of the emitting region are
shown in Figure
\ref{fg:jetexp} for all spectral epochs. The open square refers, in Figure 
\ref{fg:jetexp}, to the back-extrapolated ejection epoch of B3 (associated
with the passage of B3 through the VLBI core). At earlier epochs, the
emitting region was optically thick at frequencies $\nu\le 22$\,GHz,
which is reflected in its negative location with respect to the VLBI
core (implying that the emitting region was traveling between the jet
base and the VLBI core). The region was expanding almost linearly; the
resulting opening angle $\phi_{\rm j} = 2\fdg1 \pm 0\fdg1$ is in a
good agreement with the observed opening angle measured in the
immediate vicinity of the VLBI core (O96). During the expansion,
the magnetic field decreases from $\approx 0.2$\,G to $\approx 0.05$\,G, 
with the core magnetic field, $B_{t=1992.62} \approx 0.1$\,G.

With the jet parameters determined, we can now discuss the energy
balance in the compact jet of \object{0836+710}. We adopt the
formulation of Blandford and K\"onigl (1979) to estimate the total
(kinetic$+$magnetic field) power of the jet, $L_{\rm tot} =
2.0\cdot10^{48} $\,erg\,s$^{-1}$. The corresponding synchrotron
power of the jet is $L_{\rm syn} = 2.5\cdot 10^{47}
$\,erg\,s$^{-1}$. The data from von Linde et al. (1993) provide
an estimate of the energy deposited in the jet by the optical flare,
$E_{\rm opt} \approx 2\cdot 10^{47} $\,erg. The maximum observed
gamma luminosity of \object{0836+710} is $\approx 1.8\cdot 10^{48}
$\,erg\,s$^{-1}$ (Mukherjee et al. 1997), but it should be made
smaller by $10^2$--$10^3$ to account for Doppler boosting of the
gamma-radiation.  Comparing the optical energy output and gamma
luminosity to the jet radio power, we can conclude that flaring
activity is not likely to play a significant role in forming and
maintaining the jet.  On the other hand, we find that the average
isotropic luminosity of the optical flare is similar to the isotropic
radio luminosity observed at the epoch of the flare ($\approx 4\cdot
10^{45}$\,erg\,s$^{-1}$), suggesting the relation between the emission
in the two bands. In our view, this supports the schemes in which
relativistic electrons (responsible for synchrotron emission in the
radio and optical bands) are the primary radiating particles in the jet, and
external photons (producing the observed gamma-emission) are Compton
scattered by the jet electrons (e.g. Blandford \& Levinson 1995). Then
the energy densities of the synchrotron and Compton
emission can be related through the Doppler factor, so that $u_{\rm
C}/u_{\rm syn} = k_e \delta_{\rm eq}^{1-\alpha_{\rm C}}$, (Dermer,
Sturner
\& Schlickeiser 1997). We assume equipartition ($k_e = 1$) between the external
radiation field and the magnetic field of the emitting region. Then, for 
a typical spectral index of Compton
emission, $\alpha_{\rm C}=-1.0$, we obtain $\delta_{\rm eq} =
15.3^{+1.5}_{-1.1}$, which is consistent with the $\delta_{\rm
j}=16.4$ derived in section \ref{sc:travel}.

\section{Summary}

High frequency VLBI-observations performed shortly after a strong
gamma/optical/radio outburst in the quasar \object{0836+710} revealed the
ejection of a new jet component, which is moving at
a constant apparent speed of $\beta \sim 10$.
Back--extrapolation of the motion yields a
time of ejection of the component shortly ($< 0.7$\,yrs) after the time of the 
gamma-/optical flaring. The monitoring of the variations in total radio
spectrum showed a synchrotron self-absorbed spectral component propagating from 
high to low frequencies. The path of the spectral turnover of this component,
which we identify with the ejected jet component, is
consistent with adiabatic expansion at a slope of $S_{\rm m} \propto \nu_{\rm m}^{0.6}$.
The time lag between the outburst and the
observed component ejection can be quantitatively explained by 
strong synchrotron self--absorption at the jet base,
which prevents the detection of the radio emission from the
component at early epochs. In this model, the jet radio-core is separated
by $\approx 15$\,pc from the jet base. The jet plasma has a bulk Lorentz factor
 $\gamma_{\rm j}\approx12$, and moves at an angle of about 3$\deg$ to the line 
of sight. The corresponding jet opening angle derived from the expansion
of the spectral component is $\phi_{\rm j} = 2\fdg1$.
The derived values are consistent with the jet kinematic parameters obtained
from VLBI monitoring of the source (O96).

The high gamma-ray luminosity (Figure \ref{fg:nufnu}) of \object{0836+710},
which surpasses the IR/optical luminosity by at least one order of magnitude,
the rapid and apparently correlated variability seen in the gamma-ray and the
optical regimes (Figure \ref{fg:lightcurves}), and the ejection of a highly
superluminal ($\beta_{\rm app}=10$) jet component after the quasi-simultaneous
optical/gamma-ray flaring, strongly supports the idea 
that the gamma-emission originates in the inner parts of a
highly relativistic jet and that a broad-band correlation --
perhaps extending over the full electromagnetic spectrum --
between high energy outbursts and ejection of new jet components exists.

The observed multi band energy output during and after the flare in 1992.1
allows us to conclude that the gamma and optical flaring activity
are not sufficient to maintain the jet total power, but are connected 
with the observed jet radio emission. In this scheme, relativistic electrons
are responsible for synchrotron emission in the optical and radio bands,
and external Compton scattered photons produce the observed gamma radiation.

{\it
Acknowledgements:
We thank A.\ Marscher and A.\ Patnaik  for providing data prior to 
publication and H.\ Ter\"asranta and E.\ Valtaoja for their 
flux density measurements at 22 \& 37 GHz. We also thank H.\ Bock for
analysis of parts of the optical data and J.\ von Linde for help 
calibrating the optical data. We thank I.\ Pauliny-Toth for careful 
reading of the manuscript.
The work of T.\ P.\ K.\ was supported in part from a grant of the German
Verbundforschung of the BMBF. S.\ J.\ W.\ was supported by a grant of the
DFG (SFB 328). 

The VLBA is a facility of the NRAO, which is operated by Associated 
Universities Inc., under cooperative agreement with the NSF.

This research has made use of data from the University of 
Michigan Radio Astronomy Observatory which is supported by the National 
Science Foundation and by funds from the University of Michigan.
}

\end{document}